\documentclass[%
 reprint,
 amsmath,amssymb,
 aps,
]{revtex4-2}
\usepackage{xspace}
\usepackage{amsmath}
\usepackage{hyperref}
\usepackage{amsfonts}
\usepackage[ruled,vlined]{algorithm2e}
\usepackage{physics}
\usepackage{graphicx}
\usepackage[caption=false]{subfig}
\usepackage{booktabs}
\usepackage{xcolor}
\usepackage{orcidlink}
\usepackage{dcolumn}
\usepackage{bm}

\newcommand{\tail}{\textsc{Tailed-Uniform}\xspace}

\newcommand{\uniform}{\textsc{Uniform}\xspace}
\newcommand{\gtail}{\textsc{Gaussian-tailed}\xspace}
\newcommand{\ltail}{\textsc{Linear-tailed}\xspace}
\newcommand{\etail}{\textsc{Exponential-tailed}\xspace}
\newcommand{\ftail}{\textsc{Flat-tailed}\xspace}

\begin{document}

\title{Don't Cut Corners: How Training Outside the Prior Makes Simulation-Based Inference More Robust}

\author{Chaipat Tirapongprasert\orcidlink{0009-0000-9157-7291}}
\email{t.chaipat@columbia.edu}
\author{Matthew Ho\orcidlink{0000-0003-3207-8868}}%
\affiliation{Department of Astronomy, Columbia University, New York, NY, 10027
}

\date{\today}

\begin{abstract}
Large astrophysical simulation campaigns often generate training data by sampling parameters across a \uniform prior box. Due to the proposal's sharp edge, neural posterior estimators struggle to learn accurate approximations near the boundaries.
We propose \tail, a family of hybrid proposal distributions for sampling training simulations for robust simulation-based inference. By padding the original hard-truncated training box with decaying tails, \tail-trained networks yield more accurate posteriors near and beyond the edges. We demonstrate these improvements on a family of tail shapes, including a widened \uniform box as a control. Our results suggest that additional simulations near the prior boundary better constrain the networks as it approaches the edge of the training box, even for \uniform assumed priors. We show these advantages on a toy problem and cosmological parameter inference from the matter power spectrum. These benefits increase in high dimensions, where boundaries dominate parameter space volume.
All code is publicly available on Github.\footnote{\url{https://github.com/chaipattira/tailed-uniform-sbi.}}
\end{abstract}

\maketitle

\section{Introduction} \label{sec:intro}

Extracting physics from astronomical data is an inverse problem. We posit a model with free parameters describing the phenomenon of interest, and then ask which of those parameters could plausibly have produced the observations in hand. Traditionally, parameter estimation relied on Markov Chain Monte Carlo \citep[MCMC;][]{metropolis1953,hastings1970}, which calls for an explicit likelihood evaluation. The problem is that most nonlinear phenomena, such as the formation of large-scale structures \citep{pakmor2023millenniumtng} and binary black hole evolution \citep{siwek2023orbital}, necessitate high-fidelity forward models whose likelihoods are intractable.

Simulation-based inference \citep[SBI;][]{cranmer2020} has emerged as a framework for doing Bayesian inference on these complex simulators and is now a cornerstone methodology in cosmology \citep{alsing2019}, gravitational-wave astronomy \citep{dax2021}, and particle physics \citep{brehmer2020}.
In recent years, a number of neural-network-based SBI techniques have been proposed, such as neural posterior estimation (NPE), neural likelihood estimation, and neural ratio estimation \citep{papamakarios2018,cranmer2020,lueckmann2017}.
These approaches differ in terms of which quantity the neural network approximates (posterior, likelihood, or likelihood ratio).
For this paper, we focus on NPE, as it is the most widely adopted methodology for astrophysical problems \citep{vasist2023,dax2021,crisostomi2023}.

NPE trains on simulated $(\boldsymbol\theta, \mathbf{x})$ pairs to directly approximate the posterior, amortizing the simulation cost so that a trained network can infer posteriors for any new observation.
The reliability of these amortized posteriors depends on having a diverse and representative training set, which begins with the choice of proposal distribution from which simulation parameters are drawn.
Often, simulators preparing datasets for public simulation suites, such as the CAMELS \citep{villaescusa-navarro2021}, Quijote \citep{villaescusanavarro2020quijote}, and DREAMS \citep{rose2025introducing}, choose to simulate at a uniform distribution of points in parameter space via a Latin Hypercube bounded within a specific region of interest \citep{mckay2000}.

While efficient for volume coverage, this sampling strategy introduces a sharp discontinuity in the proposal density at the boundaries, which is difficult for gradient-descent-driven neural density estimators to model \citep{cornish2020relaxing}.
Because of the sparsity of training data, the resulting posterior becomes poorly constrained near (and beyond) the boundary of the training proposal.
This boundary pathology is exacerbated when (a) the posterior is high-dimensional, (b) the posterior has a complex shape near the boundary, and (c) a significant fraction of the posterior's probability density is outside the proposal region.

To address this problem, we propose \tail, a family of hybrid, data-efficient proposal distributions that combines \uniform cores with various boundary-extrapolating tails. We find that neural networks trained with \tail generally achieve superior posterior quality near boundaries while maintaining robust performance across the full parameter space even when a \uniform prior is assumed. In Section~\ref{sec:methods}, we define the \tail proposal.
In Section~\ref{sec:toy}, we evaluate \tail against the standard \uniform baseline on two synthetic benchmark tasks. In Section~\ref{subsec:bias}, we test the sensitivity of this advantage to tail width, training set size, network architecture, and dimensionality. In Section~\ref{sec:sci}, we apply \tail to cosmological parameter inference from the matter power spectrum. Lastly, we provide a summary of our findings in Section~\ref{sec:conclusion}.

\section{Methods} \label{sec:methods}
We define a simulator as a function $\mathcal{M}$ that maps parameters $\boldsymbol{\theta} \in \Theta \subseteq \mathbb{R}^d$ to observables $\mathbf{x} \in \mathcal{X} \subseteq \mathbb{R}^D$ \citep{cranmer2020}. Our goal is to train a neural density estimator $q_w(\boldsymbol\theta\mid\mathbf x)$ that approximates the posterior $\mathcal P(\boldsymbol\theta\mid\mathbf x)$
by minimizing
\begin{equation}
\mathcal L_{\text{NPE}} = -\mathbb E_{\mathcal D_{\text{train}}}\left[\frac{\mathcal P(\boldsymbol\theta)}{\tilde{\mathcal P}(\boldsymbol\theta)}\log q_w(\boldsymbol\theta\mid\mathbf x)\right],
\label{eq:npe-loss}
\end{equation}
where $\tilde{\mathcal{P}}(\boldsymbol\theta)$ is the proposal prior and $\mathcal{P}(\boldsymbol\theta)$ is the assumed prior. Here,
the expectation is over training pairs $\{(\boldsymbol\theta_i,\mathbf x_i)\}_{i=1}^N\sim\mathcal D_{\text{train}}$, which are generated by sampling from a proposal distribution and running the forward model \citep{papamakarios2018}. Because training pairs are drawn from $\tilde{\mathcal P}(\boldsymbol\theta)$ rather than $\mathcal P(\boldsymbol\theta)$, the importance weight $\mathcal P(\boldsymbol\theta)/\tilde{\mathcal P}(\boldsymbol\theta)$ corrects for the mismatch between the two, so that $q_w$ targets the posterior under the assumed prior regardless of how training parameters were sampled (see Section~\ref{subsec:prior} for more detail).

\begin{figure}
    \centering
    \includegraphics[width=\linewidth]{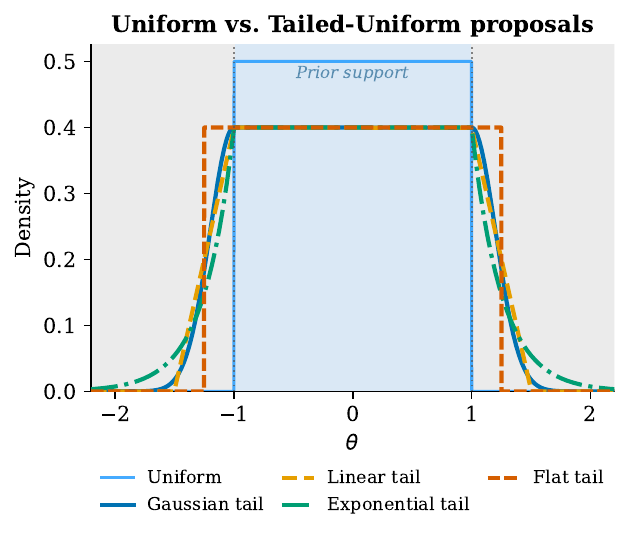}
    \caption{\tail, evaluated at $a=-1$, $b=1$, $\sigma=0.2$ ($10\%$ of the box width $W=2$), overlaid on the hard-truncated \uniform baseline. The core region is shaded blue; the extrapolation region is shaded gray.}
    \label{fig:all-proposals}
\end{figure}

\subsection{Priors}  \label{subsec:prior}

We distinguish between proposal and assumed priors.

\begin{itemize}
    \item The proposal prior $\tilde{\mathcal P}(\boldsymbol\theta)$ is the empirical distribution of parameters used to generate $\mathcal D_{\text{train}}$. It is determined by our sampling strategy.
    \item The assumed prior $\mathcal{P}(\boldsymbol\theta)$ encodes our preconceived notion of the global parameter distribution. It reflects scientific understanding, theoretical constraints, or previous empirical knowledge about plausible parameter values. Section \ref{sec:toy} will use two benchmark tasks: one with a continuous Gaussian assumed prior and the other with a uniform assumed prior hard-truncated at the edge of the proposal.
\end{itemize}

When we sample training simulations from a \uniform distribution $\tilde{\mathcal{P}} = \mathcal{U}([\theta_{\min}, \theta_{\max}]^d)$, the density drops sharply to zero at the prior boundaries. Because the network can only learn densities when the training data is well-sampled and not sparse, posteriors near boundaries are systematically unreliable.

\subsection{\tail Proposal}
To extend proposal support past the region of interest, we replace the hard-truncated \uniform proposal with a distribution that maintains a flat density on the core region but extends beyond it. Depending on the choice of profile, this construction either smooths the density transition at the original edge or simply relocates it outward.
In one dimension ($d = 1$), every member of \tail takes the form
\begin{equation}
\tilde{\mathcal P}_{\text{tail}}(x; a,b,\sigma) =
\begin{cases}
\tfrac{A}{2}\,h(a-x), & x<a\\
B\cdot\mathcal U(a,b), & x\in[a,b]\\
\tfrac{A}{2}\,h(x-b), & x>b,
\end{cases}
\label{eq:general-tail}
\end{equation}
where $a$ and $b$ define the boundaries of the \uniform core region, and $h(\cdot)$ is a unit-normalized decay profile shared by both tails. We choose $\sigma$ as a shared calibration scale, representing the half-normal width that results in tail probability mass $A$. We also impose continuity at the boundary to guarantee that our distribution is continuous and well-defined, which in turn establishes the values of the normalization constants as $A = \frac{\sqrt{2\pi}\,\sigma}{\sqrt{2\pi}\,\sigma + W}$ and $B = \frac{W}{\sqrt{2\pi}\,\sigma + W}$, where $W = b-a$ is the width of the core.

Table~\ref{tab:tail-shapes} lists the four tail shapes we compare in this paper. Widening the original \uniform proposal, \ftail is the simplest brute-force approach to addressing boundary issues.
Other profiles vary from the smooth, unbounded \gtail and \etail to the compactly supported \ltail. For all proposed distributions, we recover the hard-truncated \uniform baseline as $\sigma\to0$ (and  hence $A(\sigma)\to0$).

\begin{table}
\centering
\caption{The \tail family. $x_0$ denotes distance from the nearest core edge; each profile is continuous with the core density $B/W$ at $x_0=0$. All four are calibrated to place the same total probability mass $A(\sigma)$ beyond the core, split evenly between the two sides.}
\label{tab:tail-shapes}
\begin{tabular}{lcc}
\toprule
\textbf{Shape} & \textbf{Profile $h(x_0)$} & \textbf{Support} \\
\midrule
\gtail & $\sqrt{\tfrac{2}{\pi}}\,\tfrac{1}{\sigma} \exp\!\left(-\frac{x_0^2}{2\sigma^2}\right)$ & $[0,\infty)$ \\
\etail & $\sqrt{\tfrac{2}{\pi}}\,\tfrac{1}{\sigma} \exp\!\left(-\sqrt{\tfrac{2}{\pi}}\,\tfrac{x_0}{\sigma}\right)$ & $[0,\infty)$ \\
\ltail & $\tfrac{2}{\sqrt{2\pi}\,\sigma}\left(1-\tfrac{x_0}{\sqrt{2\pi}\,\sigma}\right)$ & $[0,\sqrt{2\pi}\,\sigma]$ \\
\ftail & $\sqrt{\tfrac{2}{\pi}}\,\tfrac{1}{\sigma}$ & $[0,\sqrt{\pi/2}\,\sigma]$ \\
\bottomrule
\end{tabular}
\end{table}

Extending to a multivariate parameter space $\boldsymbol\theta\in\mathbb R^d$, we take independent marginals of the same shape in each dimension,
\begin{equation}
\tilde{\mathcal P}_{\text{tail}}(\boldsymbol\theta; \mathbf a,\mathbf b,\boldsymbol\sigma) = \prod_{i=1}^d \tilde{\mathcal P}_{\text{tail}}(\theta_i; a_i,b_i,\sigma_i),
\end{equation}
with $\sigma_i$ set to a common fraction of the box width $b_i-a_i$ in each dimension.

\subsection{Curse of Dimensionality}\label{subsec:curse}

Given unlimited training data, one could arbitrarily extend the \uniform box until it safely encompasses every test point and posterior sample of interest. We will evaluate the feasibility of this strategy using our \ftail variant.
The concern is that at a fixed simulation budget, widening expends an increasing portion of the budget on a boundary shell that grows in volume exponentially with the number of dimensions \citep{peng2025interpretingcursedimensionalitydistance}.
For a $d$-dimensional hypercube $[a,b]^d$ with $W=b-a$, the probability that a uniformly sampled point lies within $\varepsilon$ of any face grows as $P_{\text{near bdry}}^{\text{\uniform}}(d, \varepsilon) = 1 - \left(1 - \frac{2\varepsilon}{W}\right)^d$.
\tail extends samples beyond the hypercube boundaries into tail regions to pad the training data near these boundaries. Here, we discuss the trade-off of extending this padding against reducing sample density in the internal core.

\begin{figure}
    \centering
    \includegraphics[width=1\linewidth]{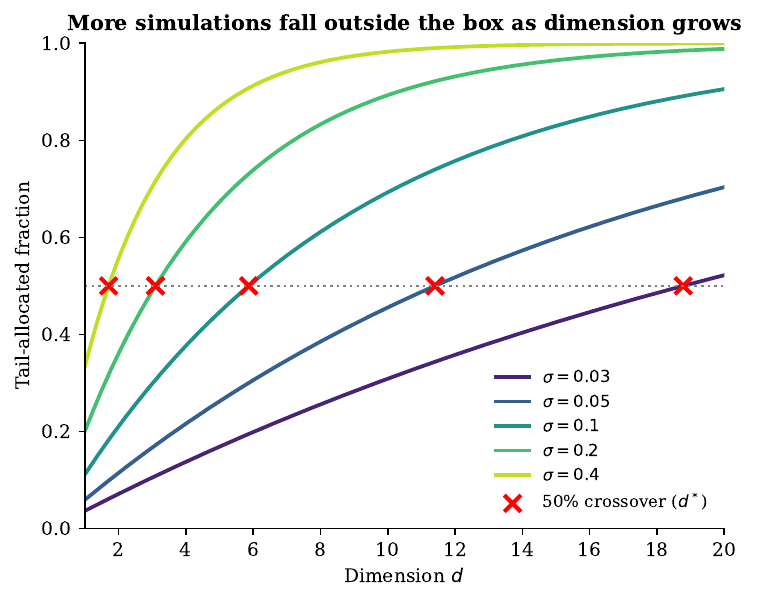}
\caption{Tail-allocated fraction $P_\mathrm{tail}^{\text{\tail}}(d) = 1-B(\sigma)^d$ versus dimension, swept across tail width $\sigma\in\{0.03,0.05,0.1,0.2,0.4\}$ ($W=2$). Markers ($\times$) show the crossover dimension $d^\ast(\sigma)$ where half the budget lands outside the box.}
    \label{fig:dim_scaling_sigma}
\end{figure}

With independent tailed marginals, the probability that a sample lands inside the $d$-dimensional hypercube $[a,b]^d$ is the product
\begin{equation}
    P_{\text{cube}}^{\tail}(d) = \prod_{i=1}^{d} B_i = B(\sigma)^d = \left(1+\frac{\sigma\sqrt{2\pi}}{W}\right)^{-d},
\end{equation}
which gives the probability of sampling in the tail regions as $P_{\text{tail}}^{\tail}(d) = 1-B(\sigma)^d$.
The selection of these tails thus reflects a bias-variance tradeoff, where we want to concentrate additional coverage precisely where it is most useful (i.e., close to the original boundary). Figure~\ref{fig:dim_scaling_sigma} shows a crossover dimension $d^\ast(\sigma)$, beyond which the tail regions hold more than half of the sampling budget.

As a heuristic, we provide the maximum tail width for a working dimension $d$ and a tolerance $p\in(0,1)$ for how much of the total budget may land in the tail:
\begin{equation} \label{eqn:sigma}
\sigma_{\max}(d;p) = \frac{W}{\sqrt{2\pi}}\left[(1-p)^{-1/d} - 1\right].
\end{equation}
While the ideal $\sigma$ appears to decrease with dimension, we show in Section ~\ref{subsec:bias} that \tail continues to function well even with a very small number (one-fifth of the total budget) of simulations within the box.

\subsection{Evaluation Protocol} \label{subsec:evaluation}

In subsequent experiments, we compare the quality of posterior estimation when training SBI models with a classic \uniform prior versus those trained with \tail priors. We measure posterior quality using the Classifier Two-Sample Test \citep[C2ST;][]{lopez-paz2018}, which was shown to be the most sensitive probe of local posterior quality among the metrics compared in a large-scale SBI benchmarking study \citep{lueckmann2021}. At an observation $\mathbf x_0$, we train an independent classifier to distinguish samples from two distributions: a reference posterior (analytic, where available, or MCMC otherwise) and a surrogate network $q_w(\boldsymbol\theta\mid\mathbf x_0)$ trained on some proposal prior. We report the classifier's held-out misclassification rate: a score close to $0.5$ indicates that the classifier struggles to differentiate between the two sample sets, implying that the distributions are indistinguishable (i.e. good performance), whereas a score close to $0$ indicates maximally distinguishable distributions (i.e. poor performance).

\begin{figure*}
    \centering
    \subfloat[Gaussian Linear Task]{\includegraphics[width=0.5\linewidth]{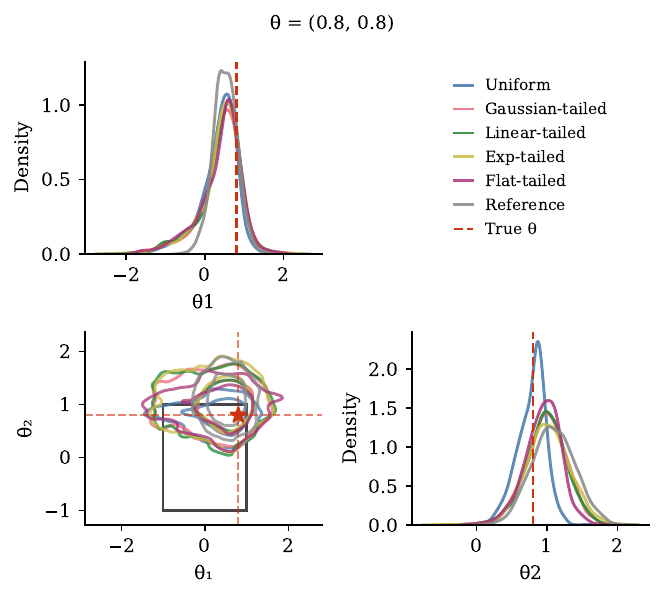}}
    \hfill
    \subfloat[Gaussian Linear Uniform Task]{\includegraphics[width=0.5\linewidth]{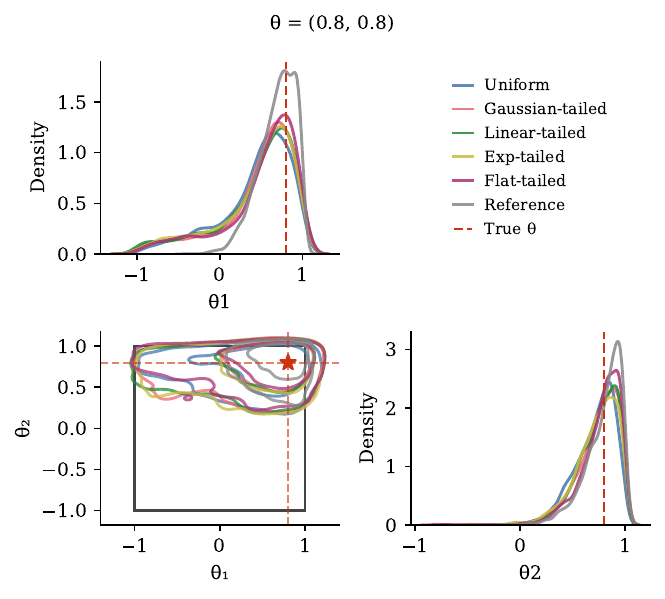}}
    \caption{Corner plots comparing posterior estimation performance for the boundary test case $\boldsymbol\theta_{\text{true}}=(0.8,0.8)$. \textbf{(a)} Under a Gaussian assumed prior, the reference posterior (gray) extends past the training box $[-1,1]^2$ (black square). \textbf{(b)} Under a Uniform assumed prior, the reference is constrained to lie within the box by construction. In both panels, \uniform (blue) is the most poorly constrained; all four \tail variants (warm colors) track the reference more closely, despite the two references having qualitatively different shapes near the boundary.}
    \label{fig:onetest}
\end{figure*}

To characterize how posterior quality depends on proximity to proposal prior boundaries, we evaluate C2ST at various regions of parameter space by generating test points on a regular $10\times10$ grid spanning the training box. We also average the resulting C2ST scores radially, from the interior to the boundary and the extrapolation region beyond.
This distance-binned protocol is used across all proposals and tail shapes examined in this paper.

\section{Toy Problem} \label{sec:toy}
We first demonstrate \tail on two benchmark tasks where the true posterior is analytically tractable. Taken from the SBI benchmark suite \citep[\texttt{sbibm};][]{lueckmann2021}, both tasks share a $d$-dimensional Gaussian simulator $\mathcal{M}(\boldsymbol{\theta}) = \boldsymbol{\theta}$ on $\boldsymbol{\theta} \in [-1,1]^d$ and a Gaussian likelihood $\mathcal P(\mathbf{x}\mid\boldsymbol{\theta}) = \mathcal{N}(\mathbf{x}; \boldsymbol{\theta}, \mathbf{I}_d)$. The only difference is in the assumed prior $\mathcal P(\boldsymbol\theta)$, as alluded to in Section~\ref{subsec:prior}:
\begin{itemize}
    \item \emph{Gaussian Linear Task} uses a standard Gaussian assumed prior $\mathcal P(\boldsymbol\theta) = \mathcal N(\mathbf 0, \mathbf I_d)$, allowing the posterior to leak beyond the training box $[-1,1]^d$.
    \item \emph{Gaussian Linear Uniform Task} uses a uniform assumed prior hard-truncated at $[-1,1]^d$. Because its coverage corresponds to the proposal, the posterior is guaranteed to vanish outside of the boundary of the training box.
\end{itemize}

We aim to compare the performance of five NPEs, one trained with the original $\tilde{\mathcal P}_{\text{\uniform}} = \mathcal U([-1,1]^d)$ and the other four with our \tail proposals (with $\sigma_i = 0.1\times(b_i - a_i) = 0.2$).

\subsection{Training}
For each proposal, we generate $N = 3{,}000$ simulation pairs $\{(\boldsymbol\theta_i, \mathbf x_i)\}_{i=1}^N$ via Latin hypercube sampling in the proposal's quantile space, run the forward simulator $\mathbf x_i = \mathcal M(\boldsymbol\theta_i) + \boldsymbol\epsilon_i$ with $\boldsymbol\epsilon\sim\mathcal N(\mathbf 0,\mathbf I_d)$, and train an NPE ensemble on each resulting dataset.
Our default ensembles consist of a network of Masked Autoregressive Flow \citep[MAF;][]{papamakarios2018maf} and Masked Autoencoder for Distribution Estimation \citep[MADE;][]{germain2015} density estimators, each with 50 hidden features and 5 autoregressive transform layers, trained with batch size 64 and learning rate $5\times10^{-5}$.
The pipeline is built on the \texttt{LtU-ILI} (Learning the Universe—Implicit Likelihood Inference) framework \citep{ho2024}, which offers standardized data handling and ensemble utilities for simulation-based inference. 

Because both tasks are analytically tractable and low-dimensional, the posteriors are simple enough that any reasonable configurations converge to acceptable solutions. Section~\ref{subsec:bias} varies one axis at a time to test the sensitivity of our results to tail width, simulation budget, network architectures, and dimensionality.

\begin{figure*}
    \centering
{\includegraphics[width=\linewidth]{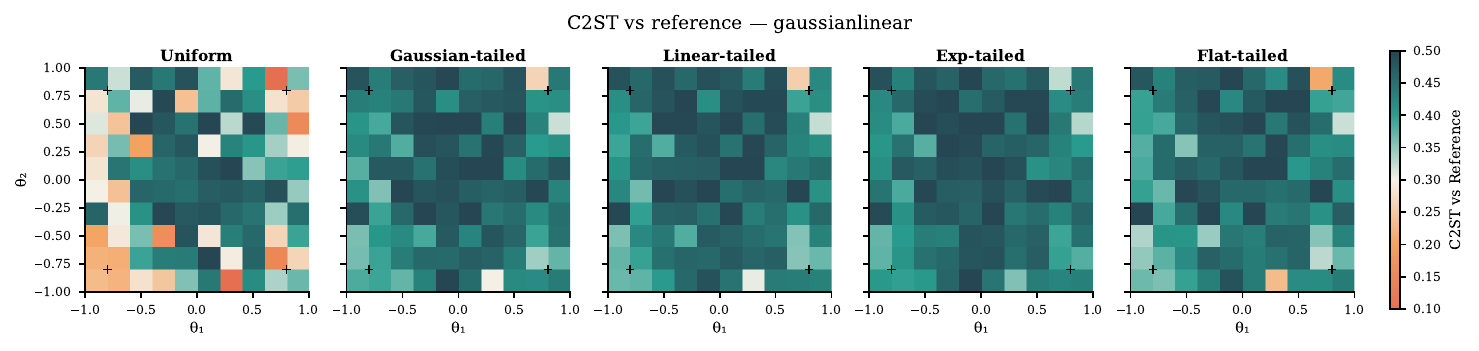}}\\
{\includegraphics[width=\linewidth]{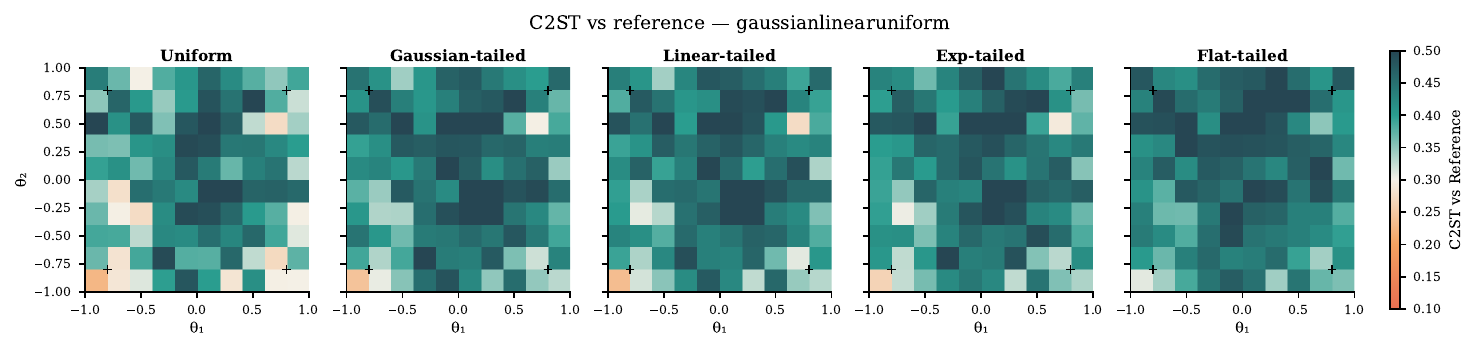}}
    \caption{C2ST against the reference posterior on the $10\times10$ grid, one panel per proposal, under each assumed prior. Orange marks poor agreement with the reference; teal/blue marks almost-indistinguishable agreement. \textit{\uniform} (leftmost column, both rows) shows orange patches concentrated near the edges and corners; every \tail NPE replaces most of these with blue at the same locations, in both tasks.}
    \label{fig:heatmap}
\end{figure*}

\subsection{Validation}

As a first pass at $d=2$, we select a test point $\boldsymbol\theta_{\text{true}} = (0.8, 0.8)$ near the boundary of parameter space, generate an observation $\mathbf x_0 = \mathcal M(\boldsymbol\theta_{\text{true}}) + \boldsymbol\epsilon$ and draw $M=2{,}000$ posterior samples from the analytic reference and from each of the five NPEs.
Our goal is to assess how well each method constrains the parameters by examining the distribution and concentration of samples around the true value. Figure~\ref{fig:onetest} and Table \ref{tab:onetable} illustrate that each variant of \tail-trained NPEs consistently outperforms \uniform-trained NPE.

\begin{table}[ht]
\centering
\caption{Posterior performance for $\boldsymbol\theta_{\text{true}} = (0.8, 0.8)$}
\label{tab:onetable}
\begin{tabular}{lccc}
\toprule
\textbf{Method} & \multicolumn{2}{c}{\textbf{Parameter Estimates}} & \textbf{C2ST} \\
\cmidrule(lr){2-3}
& $\theta_1$ & $\theta_2$ & \\
\midrule
\multicolumn{4}{l}{\textit{Gaussian assumed prior}} \\
Reference & $0.516 \pm 0.307$ & $1.092 \pm 0.304$ & 0.500 \\
\uniform & $0.351 \pm 0.469$ & $0.781 \pm 0.207$ & 0.268 \\
\gtail & $0.396 \pm 0.537$ & $0.991 \pm 0.287$ & \textbf{0.424} \\
\ltail & $0.397 \pm 0.548$ & $0.990 \pm 0.288$ & \textbf{0.441} \\
\etail & $0.403 \pm 0.528$ & $1.014 \pm 0.315$ & \textbf{0.433} \\
\ftail & $0.390 \pm 0.552$ & $0.942 \pm 0.257$ & \textbf{0.418} \\
\midrule
\multicolumn{4}{l}{\textit{Uniform assumed prior}} \\
Reference & $0.690 \pm 0.219$ & $0.799 \pm 0.169$ & 0.500 \\
\uniform & $0.433 \pm 0.431$ & $0.717 \pm 0.189$ & 0.356 \\
\gtail & $0.460 \pm 0.451$ & $0.746 \pm 0.197$ & \textbf{0.384} \\
\ltail & $0.446 \pm 0.470$ & $0.745 \pm 0.188$ & \textbf{0.382} \\
\etail & $0.442 \pm 0.463$ & $0.731 \pm 0.201$ & \textbf{0.383} \\
\ftail & $0.490 \pm 0.451$ & $0.772 \pm 0.182$ & \textbf{0.424} \\
\bottomrule
\end{tabular}
\end{table}

For the Gaussian Linear task, we see that every posterior (including the reference) leaks outside the proposal boundaries, despite being trained exclusively on simulations within this region. The observed leakage occurs because the task adopts a Gaussian assumed prior, which is nonzero everywhere. Without gradient information at the boundaries, our flow-based density estimator cannot learn the correct density transitions near the cutoff.
This under-sampling affects the \uniform-trained NPE the most, as it has never seen samples beyond the boundary, resulting in diffuse and poorly constrained samples. By allocating some budget outside the original box, \tail-trained NPEs can extrapolate to better match the reference posterior at boundary-adjacent test points.

Now, for the Gaussian Linear Uniform task, the reference posterior is hard-truncated at the boundary by the uniform assumed prior. One might assume that simulating outside in this setting is just wasteful, as far-tail samples do not contribute to the loss function. In fact, training with \tail still provides a small but consistent advantage, as seen in Table \ref{tab:onetable}. With extra samples outside the original boundaries, \tail simulations can better constrain network behavior as it approaches the boundary.

Next, we evaluate all five NPEs' performance on a uniform $10\times10$ grid spanning $[-1,1]^2$ (100 points). We draw $M=2{,}000$ posterior samples per point and compute pairwise C2ST against the reference at each. Figure~\ref{fig:heatmap} reveals that \uniform's performance deteriorates as it drifts away from the prior center, as evidenced by the proliferation of orange patches near the boundary. In contrast, our \tail maintains more teal and blue coloration (i.e., better performance) across the parameter space.

\begin{figure}
    \centering
\includegraphics[width=0.9\linewidth]{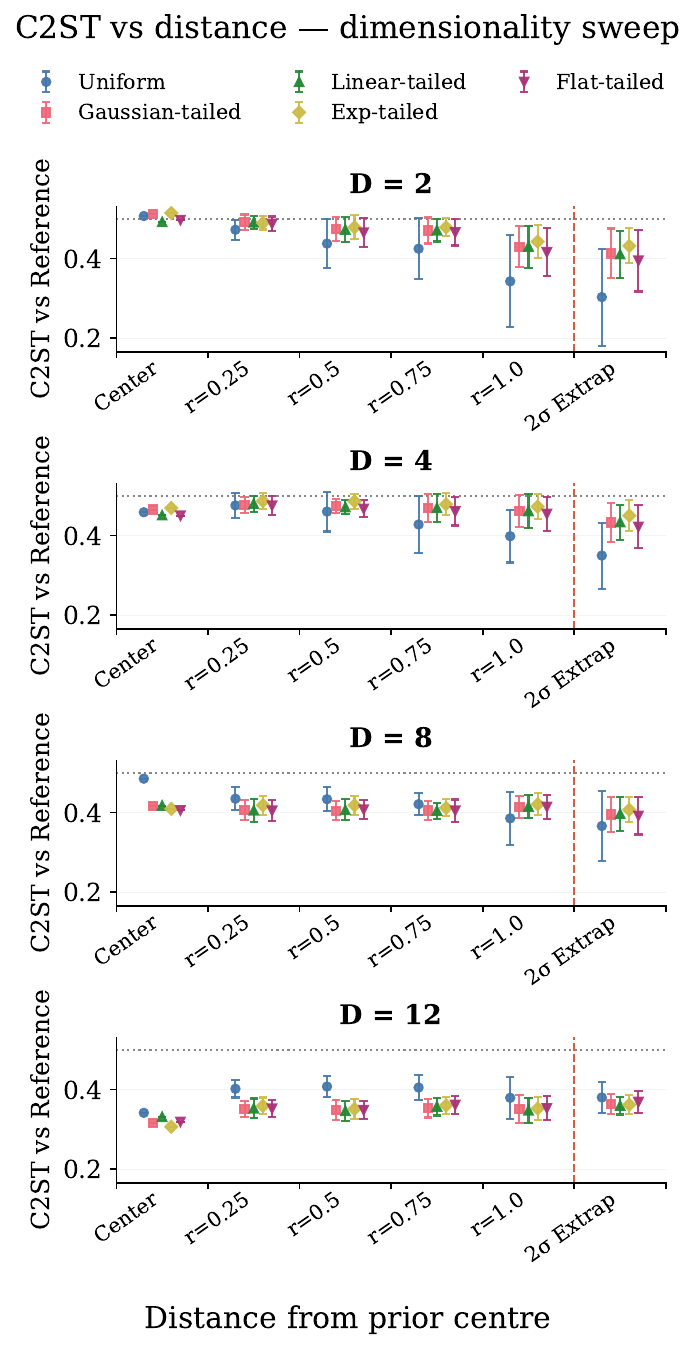}
    \caption{C2ST versus distance from the prior center across dimensions $d\in\{2,4,8,12\}$ at fixed $N=3{,}000$ and $\sigma = 0.2$ for Gaussian Linear task. Error bars: 16th--84th percentile across test points in each bin.}
    \label{fig:dim}
\end{figure}

\section{Sensitivity Analysis} \label{subsec:bias}
In this section, we investigate whether the benefit of the \tail is unique to one tail shape, how the trade-off scales with dimension, and whether more training data or more network capacity alone can mitigate the boundary pathology. All sensitivity studies below use the Gaussian assumed prior.

\subsection{Does \tail work in higher dimensions?}

The tail width $\sigma$ encodes a bias-variance trade-off: narrower tails limit how much of the boundary pathology can be resolved, while wider tails divert an increasing share of the fixed simulation budget away from the region of interest. It is reasonable to assume that diluting the samples may diminish the efficacy of \tail. To assess how the optimal $\sigma$ depends on the dimensionality, we first extend the Gaussian Linear task to higher dimensions, $d\in\{2,4,8,12\}$, and train NPEs at the default $N=3{,}000$ simulations and $\sigma=0.2$.

\begin{figure}
    \centering
    \includegraphics[width=0.9\linewidth]{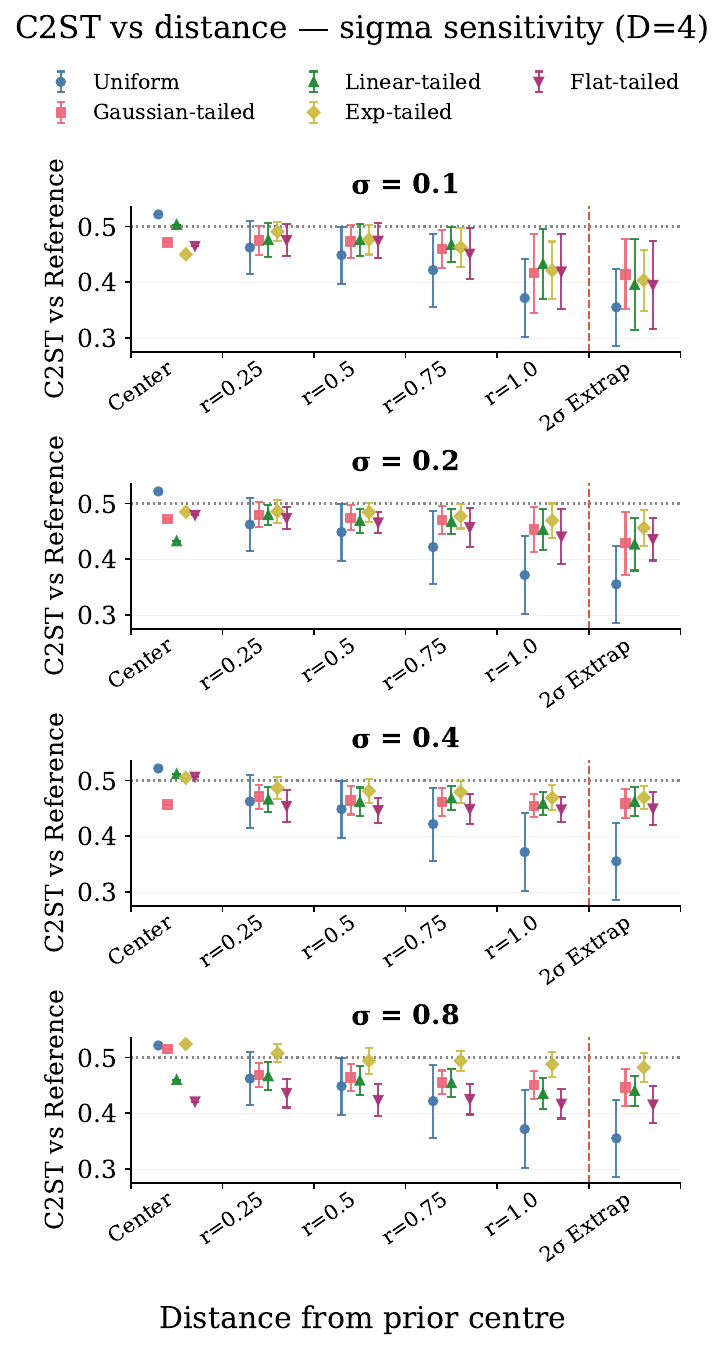}
    \caption{C2ST versus distance from the prior center, stratified by tail width $\sigma$, at $d=4$ for Gaussian Linear task. Error bars: 16th--84th percentile across test points in each bin.}
    \label{fig:ablation_sigma}
\end{figure}

Figure~\ref{fig:dim} demonstrates that \tail continues to outperform \uniform throughout the whole parameter space with over $59\%$ of its budget in the tails at $d = 4$ (see Equation~\ref{eqn:sigma}). With increasing dimensions, the tail-allocated fraction eventually consumes so much of the budget that too little remains to resolve the interior or the boundary. This tradeoff begins to manifest in $d = 8$ ($83\%$ of the budget in the tail region), where \tail still performs better near the boundary and at the extrapolation bin but begins to plummet inside. Only at $d=12$, when $93\%$ of the budget is diverted outside, does the advantage disappear.

We see a similar behavior in Figure~\ref{fig:ablation_sigma} for $d = 4$ dimensions and $N=3{,}000$ simulations. While \uniform consistently degrades as $r>0.5$, \tail achieves robust performance throughout the near-boundary and extrapolation bins.
Putting $p=50\%$ into Equation~\ref{eqn:sigma}, we can find the halfway crossover point (where the tails have received more than half of the total budget), which is at $\sigma_{\max}\approx 0.151$ for $d = 4$. Having less than half of the samples in the original central region of interest does not result in poor performance near the boundary for $\sigma > \sigma_{\max}$.

These results imply that the marginal value of the tail samples outweighs the cost of diluting interior coverage. At moderate $\sigma$, the choice of tail shape matters far less than the choice of having a tail, as shown by \ftail's consistent performance with the other tailed variants. However, at higher $\sigma$, merely sampling beyond the original proposal boundary is insufficient; the tails also need to be smooth and differentiable.

\begin{figure}
    \centering
    \includegraphics[width=\linewidth]{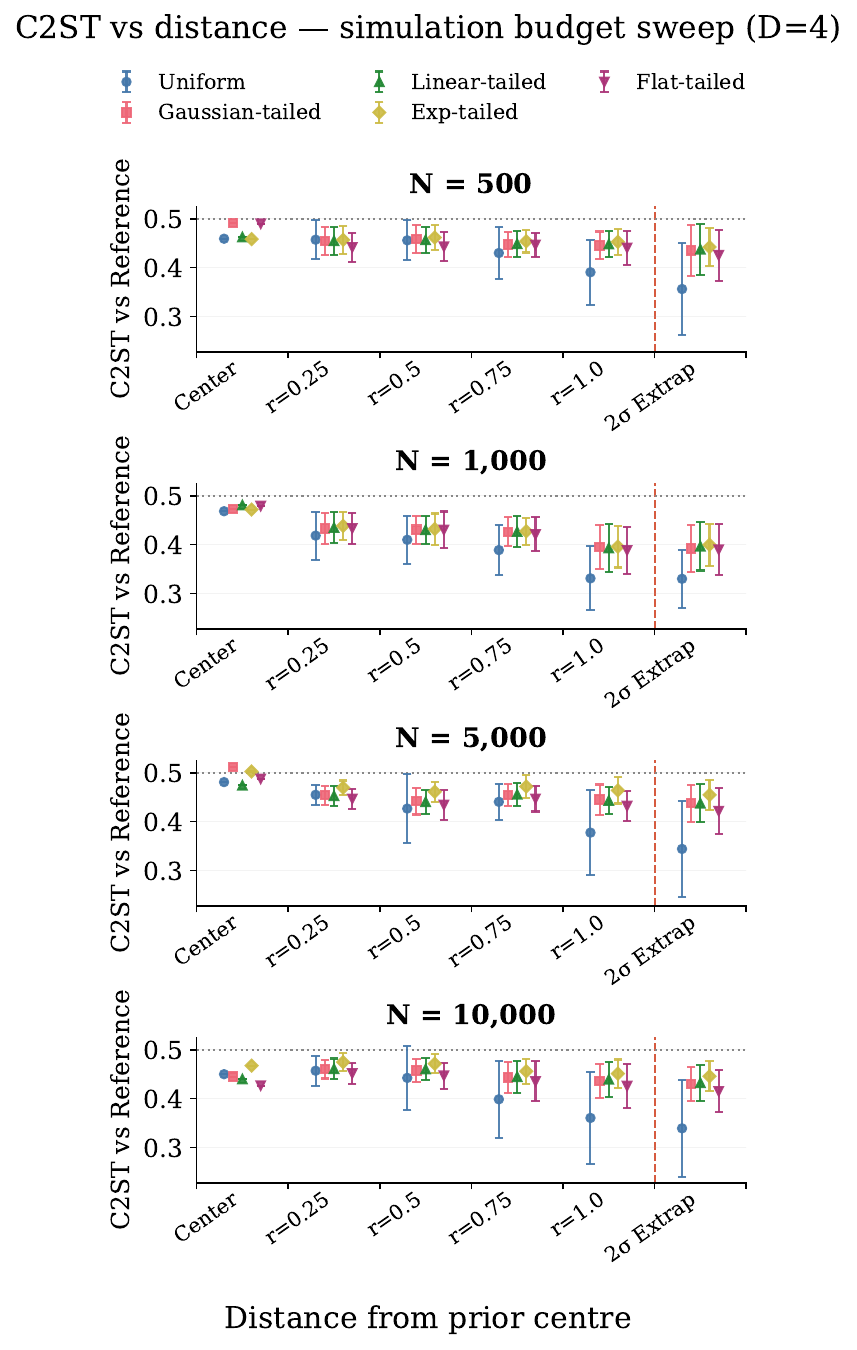}
    \caption{C2ST versus distance from the prior center, stratified by simulation budget $N$, at $d=4$  for Gaussian Linear task. Error bars: 16th--84th percentile.}
    \label{fig:ablation_nsims}
\end{figure}

\subsection{Does adding more training data help?} \label{subsec:training_data}
If only insufficient data in the near-boundary regions causes degradation, increasing the training set size will solve the problem.
We test this by varying $N \in \{500, 1000, 5000, 10000\}$ at fixed $\sigma = 0.2$ at $d=4$.

Empirically, we find in Figure~\ref{fig:ablation_nsims} that increasing the number of simulations across two orders of magnitude barely helps with the NPE's performance. Across all four panels, \uniform shows the expected degradation pattern near the boundaries, while all \tail variants achieve a consistent C2ST score above $0.42$.
This suggests that the boundary degradation stems not from absence of
data but from the extent of the proposal's support. Since orders-of-magnitude more supplementary data cannot supply information about the region beyond the \uniform's edge, one can achieve better posterior quality at a fraction of the computational cost just by sampling more judiciously.

\begin{figure}
    \centering
    \includegraphics[width=0.9\linewidth]{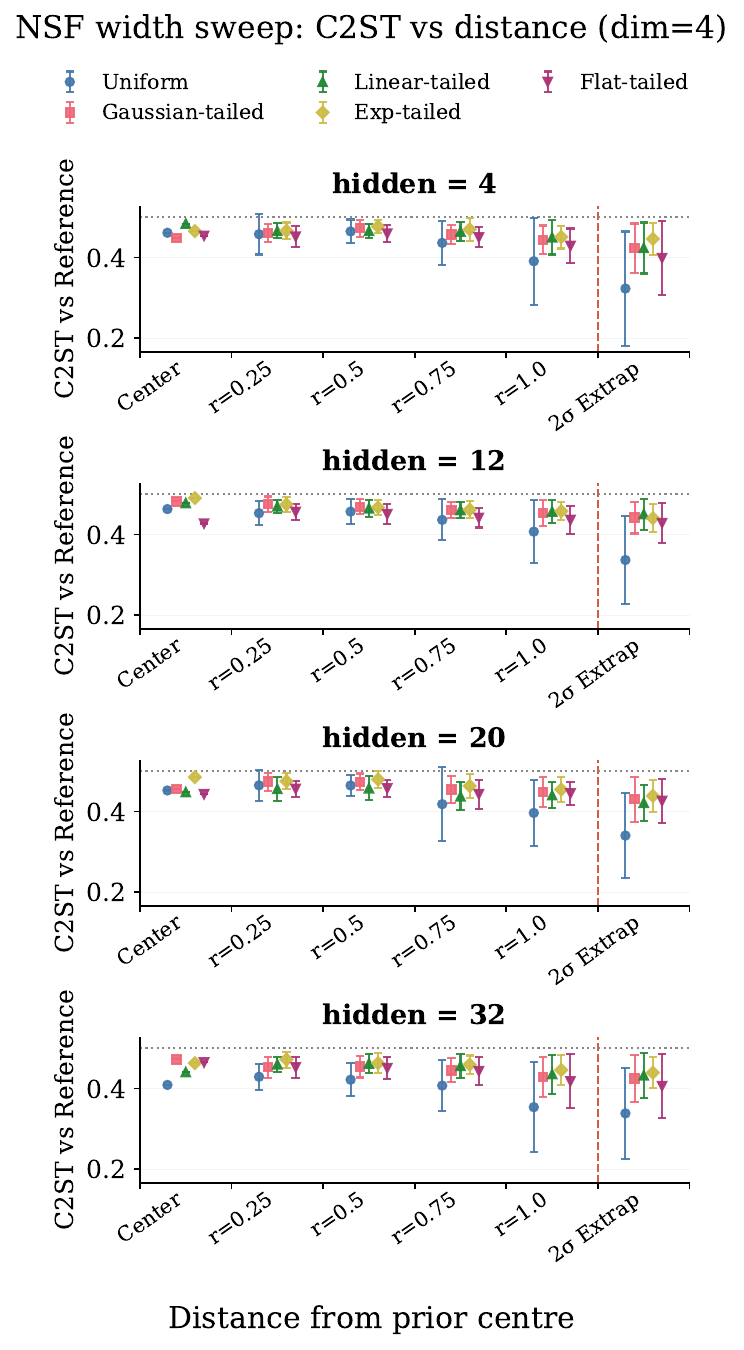}
    \caption{C2ST versus distance from the prior center across NSF hidden-layer widths for the 4-dimensional Gaussian Linear task. \tail's superiority over \uniform persists despite an eightfold increase in network width.}
    \label{fig:capacity}
\end{figure}

\subsection{Does changing the architecture help?}\label{subsec:architecture}

Another objection is that the boundary discontinuity may be unique to our default configuration, and switching to a thinner or wider architecture may resolve it.
We test this by using a single neural spline flow \citep[NSF;][]{durkan2019neuralsplineflows} at varying hidden layer widths. Since NSFs are universal density approximators, they should be able to approximate this relatively simple Gaussian posterior if the underlying problem is architectural.

Figure~\ref{fig:capacity} shows that the boundary degradation persists, and the performance across all four widths is roughly the same. Together with Section \ref{subsec:training_data}, this indicates that the pathology is a property of the proposal's support rather than an expressivity bottleneck or a data-volume limitation. No amount of additional network capacity or interior simulations can recover data that was never sampled beyond the edge.

\section{Cosmological Application} \label{sec:sci}

Following the same training and validation pipeline in Section~\ref{sec:toy}, we now apply \tail to a popular inference benchmark in cosmology: the inference of four cosmological parameters $\boldsymbol{\theta} = (\Omega_m, h, n_s, A_s)$ from the matter power spectrum $\mathbf{P} \in \mathbb{R}^{64}$, where $\Omega_m$ is the matter density parameter, $h$ is the dimensionless Hubble parameter, $n_s$ is the spectral tilt, and $A_s$ is the primordial amplitude (in units of $10^{-9}$).
\subsection{Training}
We generate observations using the \texttt{syren-new} emulator \citep{sui2025} with the additional heteroskedastic cosmic variance noise \citep{dodelson2020}.
For a comoving survey volume of $V = L^3$, the longest wavelength that fits in the box (i.e., the fundamental mode) is
$k_f = \abs{\mathbf k} = \frac{2\pi}{L}$. Assuming the universe is homogeneous and isotropic (i.e., rotationally invariant) on large scales \citep{dodelson2020}, we find the number of modes inside a spherical shell at wavenumber $k$ with thickness $\Delta k = k_f$ to be
\begin{equation}
    N_k = \frac{4\pi k^2 \Delta k}{k_f^3} = \frac{L^3k^2k_f}{2\pi^2}.
\end{equation}
Together, we can construct an observation as
\begin{equation}
\mathbf x_i = \mathbf P(k \mid \boldsymbol\theta_i) = \underbrace{\mathbf P_{\text{theory}}(k \mid \boldsymbol\theta_i)}_{\text{\texttt{syren\_new} emulator}} + \underbrace{\boldsymbol{\epsilon}_i}_{\text{cosmic noise}},
\end{equation}
where $\boldsymbol\epsilon_i\sim \mathcal{N}\left(\mathbf{0}, \text{diag}(\sigma^2_1, \ldots, \sigma^2_{N_k})\right)$ represents the heteroskedastic (scale-dependent) cosmic noise with a variance $\sigma_k^2 = \frac{2\mathbf P_{\text{theory}}^2(k \mid \boldsymbol\theta_i)}{N_k}$.

We choose a \uniform prior box as $\Omega_m \in [0.27, 0.37]$, $h \in [0.63, 0.71]$, $n_s \in [0.94,0.98]$, and $A_s \in [1.9,2.3]$. For the \tail proposals, we pick the tail width $\sigma_i = 0.1\times$(box width) in each dimension. We train all five NPEs at $N=3{,}000$ simulations with a MAF layer with $36$ hidden features and $4$ transforms. As reference posteriors, we use MCMC to collect samples under a log-normal prior on $(\Omega_m, h)$ truncated at $\pm3\sigma$.

\begin{figure}
    \centering
    \includegraphics[width=\linewidth]{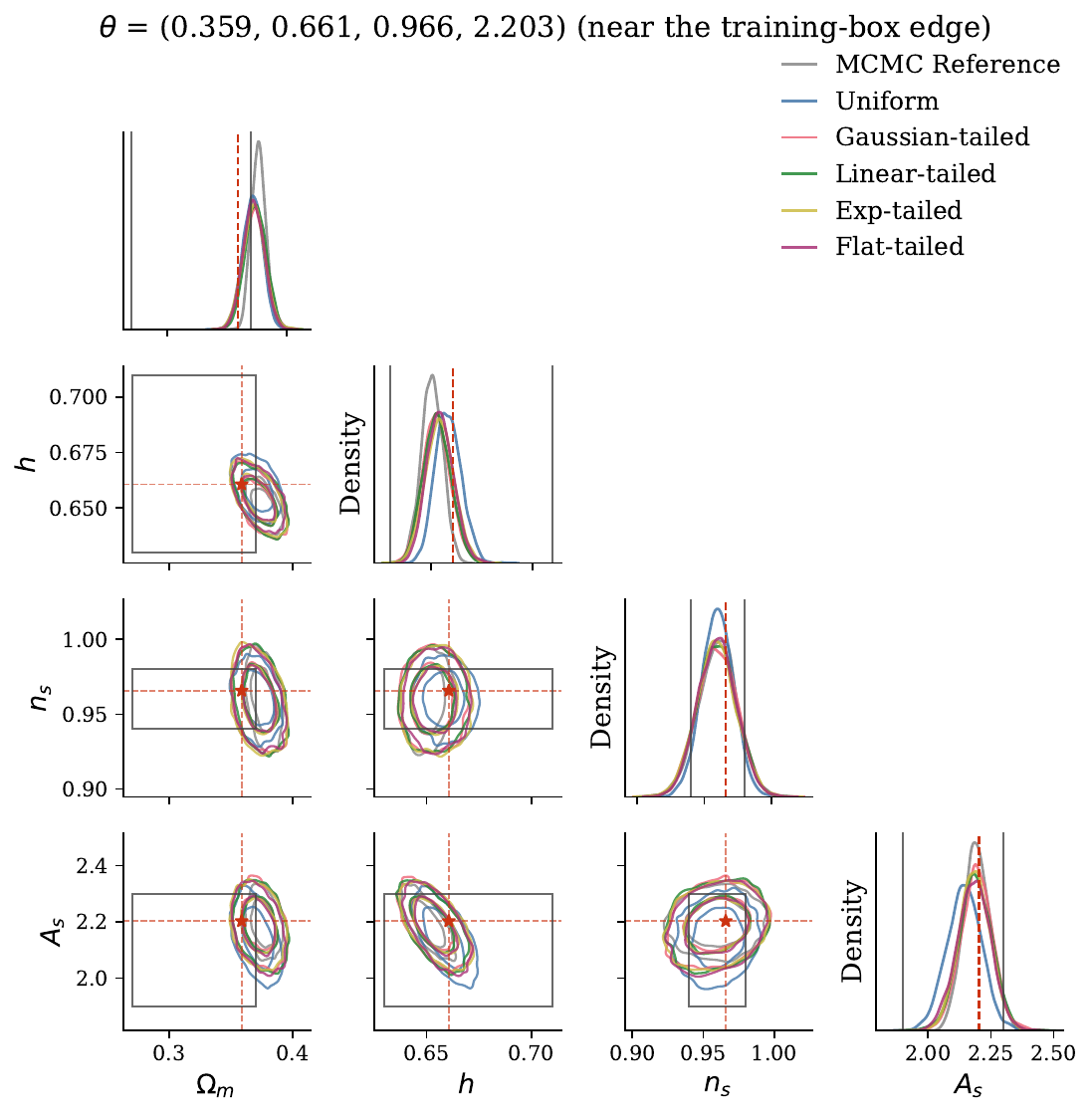}
    \caption{Corner plot comparing posterior estimation performance at $\boldsymbol\theta_\text{true} = (\Omega_m,h,n_s,A_s) = (0.359, 0.661, 0.966, 2.203)$. MCMC reference in gray; \uniform NPE in blue; the four \tail NPE variants in red (\gtail), green (\ltail), yellow (\etail), and purple (\ftail). Solid gray lines mark the edges of the original training box; the true value is marked by the red star.}
    \label{fig:corner-sci-4d}
\end{figure}

\begin{figure*}
\centering
\includegraphics[width=\linewidth]{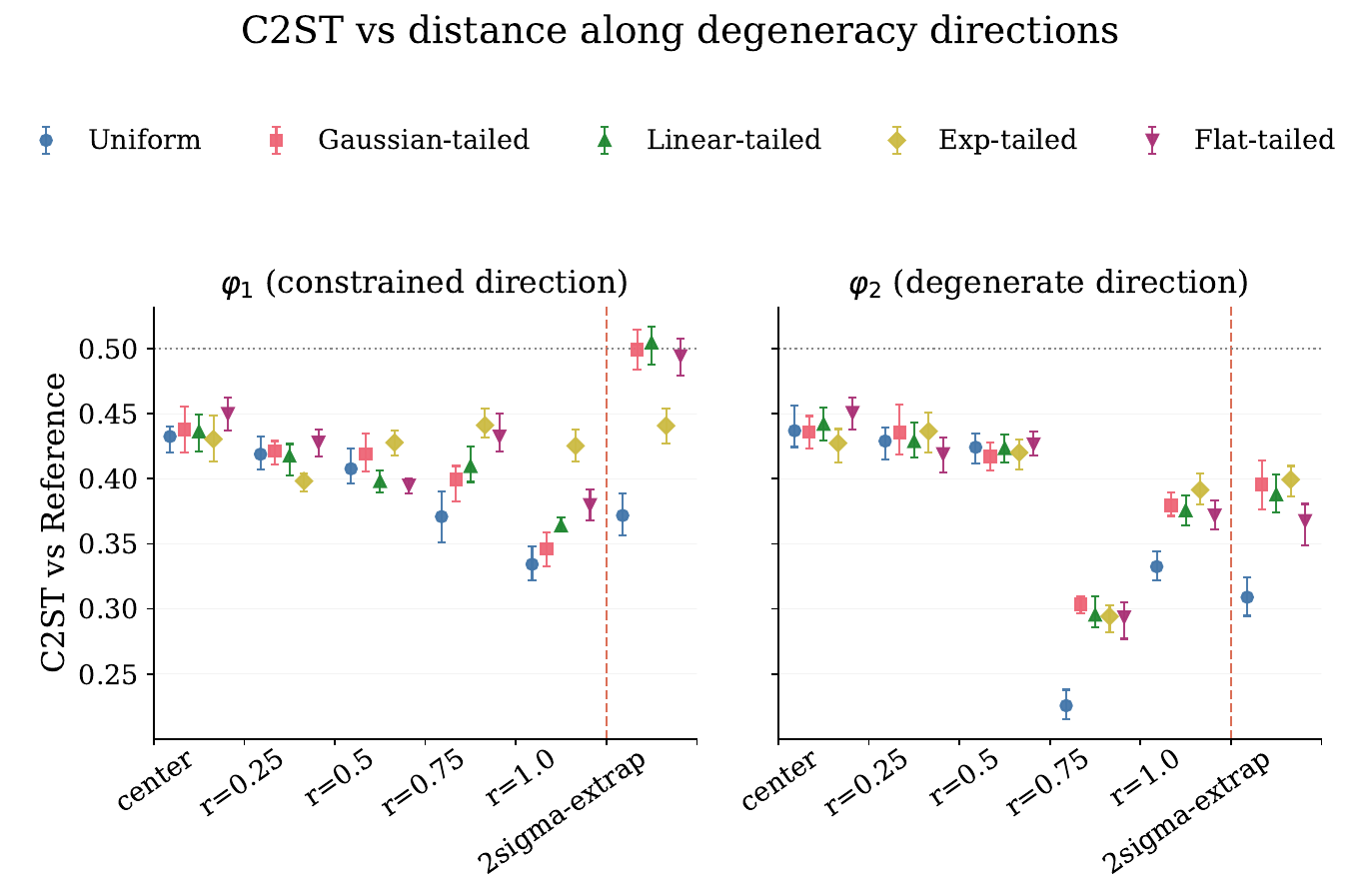}
\caption{C2ST vs.\ position along two orthogonal directions (degenerate and constrained). Error bars: 16th--84th percentile.}
\label{fig:directional-sci}
\end{figure*}

\begin{table*}
\centering
\caption{Posterior performance for test point $\boldsymbol\theta_\text{true} = (\Omega_m,h,n_s,A_s) = (0.359, 0.661, 0.966, 2.203)$.}
\label{tab:sci-testpoint-4d}
\begin{tabular}{lccccc}
\toprule
\textbf{Method} & \multicolumn{4}{c}{\textbf{Parameter Estimates}} & \textbf{C2ST} \\
\cmidrule(lr){2-5}
& $\Omega_m$ & $h$ & $n_s$ & $A_s$ & \\
\midrule
MCMC Reference & $0.376\pm0.006$ & $0.650\pm0.005$ & $0.960\pm0.014$ & $2.198\pm0.052$ & 0.500 \\
\midrule
\textit{\uniform} & $0.371\pm0.008$ & $0.659\pm0.006$ & $0.960\pm0.012$ & $2.142\pm0.067$ & 0.307 \\
\gtail & $0.374\pm0.009$ & $0.653\pm0.007$ & $0.959\pm0.014$ & $2.192\pm0.062$ & \textbf{0.458} \\
\ltail & $0.374\pm0.009$ & $0.653\pm0.007$ & $0.960\pm0.014$ & $2.189\pm0.065$ & \textbf{0.452} \\
\etail & $0.372\pm0.009$ & $0.654\pm0.007$ & $0.958\pm0.014$ & $2.187\pm0.062$ & \textbf{0.430} \\
\ftail & $0.373\pm0.009$ & $0.655\pm0.006$ & $0.960\pm0.014$ & $2.182\pm0.067$ & \textbf{0.419} \\
\bottomrule
\end{tabular}
\end{table*}

\subsection{Validation}
We adopt a similar spatial evaluation framework as in Section~\ref{sec:toy} with a few modifications. First, the \uniform training box is anisotropic in this setting, so we put distance-to-boundary on a common scale across dimensions by sampling the evaluation points under the per-dimension half-width coordinate $\mathbf z = (\boldsymbol\theta-\boldsymbol\theta_c)/\mathbf{h} \in [-1,1]^4$, where $\mathbf h$ is the box half-width vector and $\boldsymbol\theta_c$ is the box's center.
Second, the matter power spectrum is sensitive to the combination of parameters, such as $(\Omega_m, h)$, rather than each parameter independently \citep{peacock1996}. These degeneracies, as seen in the clustering of posterior samples along a slanted contour for some parameter pairs (see Figure~\ref{fig:corner-sci-4d}), could reduce the constraining power of posteriors along the degenerate direction.

To inspect this effect, we use \textsc{DegenDetector} \citep{2026arXiv260708755T} to identify the candidate degeneracies in our MCMC chain and recover their functional forms. The top three couplings are $(\Omega_m,h)$, $(h,A_s)$, and $(n_s,A_s)$. Picking the strongest ones, we recover the equation
\begin{equation}
\varphi_2 = -n_2\ln\Omega_m + n_1\ln h,
\label{eq:phi2}
\end{equation}
where $\varphi_2$ is the degeneracy direction with $n_1 = 0.363$ and $n_2 = 0.932$. Accordingly, we also define the orthogonal direction $\varphi_1 = n_1\ln\Omega_m + n_2\ln h$, where we anticipate a more constrained posterior throughout.

Figure~\ref{fig:corner-sci-4d} compares the MCMC reference against all five NPEs at $\boldsymbol\theta_\text{true} = (0.359, 0.661, 0.966, 2.203)$, which lies at the edge of the normalized prior box. Despite the anisotropic posteriors, the four \tail variants more closely match the MCMC reference, with samples more concentrated around true parameter values. Table~\ref{tab:sci-testpoint-4d} quantifies this result.

To evaluate performance across the entire parameter space,
we fix $n_s$ and $A_s$ at the box center and sweep along two different binning directions $\varphi_1$ and $\varphi_2$ in $(\Omega_m, h)$ space, drawing $M=2{,}000$ samples per posterior. Figure~\ref{fig:directional-sci} shows the five NPEs performing comparably from the center to the midway point along both $\varphi_1$ and $\varphi_2$. While \tail remains robust throughout the parameter space, \uniform's performance starts to deteriorate as $r>0.75$ through the $2\sigma$ extrapolation ring. Such degradation is more severe in the degenerate direction as expected.
In contrast to Section~\ref{sec:toy}, the overarching trend here is less clean, as evidenced by the dip at $r=0.75$ for $\varphi_2$ across all NPEs. We attribute this precipitous plunge in performance to the confounding effects of other degenerate pairs as well as statistical noise fluctuation of this particular realization.

\section{Conclusion} \label{sec:conclusion}

In this paper, we propose \tail, a family of hybrid proposal distributions that pad the sharp boundary of the standard \uniform prior with boundary-extrapolating tails. This padding mitigates the structural pathology that causes NPE posteriors to degrade near the edges of parameter space. With minimal hyperparameter tuning, \tail enables more robust inference across most of the configurations we test.

In both the synthetic and cosmological benchmarks, \tail-trained NPE samples closely resemble the reference posteriors across the full parameter space, while \uniform posteriors leak probability mass beyond the prior support and fail to constrain parameters near the boundary. Our proposal's merit also grows in higher dimensions, where an exponentially increasing fraction of parameter space lies near boundaries.

This advantage is not without drawbacks, as more training samples are diverted into the tails as dimensionality increases. However, our sensitivity analyses show that the marginal benefit of allocating simulations outside the original bounds is considerably greater than the cost of reduced internal coverage. Among all the \tail proposals, smooth profiles are more favorable due to their insensitivity to tail width choices.
Additionally, we demonstrate that the boundary degradation of \uniform-trained NPE is structural, as the \uniform proposal fails to provide sufficient sampling density in the vicinity of the edges.
Increasing training data by two orders of magnitude or network width eightfolds thus yields negligible improvement, whereas any proposal that allocates budget beyond the original bounds recovers most of the lost accuracy.

Future work could look into a variety of avenues. First, it is not clear if we should impose a hard cutoff for the tails at the physical boundary or construct proxy simulations in the unphysical regions. Second, developing a principled and rigorous approach for selecting per-parameter $\sigma_i$ values would be beneficial, especially when some parameters have different ranges.
Third, rather than sampling independently along the physical axes, one could reparametrize to smooth along the principal directions of the posterior in problems with strong degeneracies. We believe that sampling in the rotated basis might provide more coverage, resulting in improved performance. Finally, testing beyond $d=12$ and scaling $\sigma$, $N$, and $d$ together would provide a clearer picture of the simulation budget required to perform well for a given inference task.

\bibliography{apssamp}

\end{document}